\documentclass[pra,aps,twocolumn,epsfig,nofootinbib]{revtex4}
\usepackage{amsmath}
\usepackage{amssymb}
\usepackage[dvips]{graphicx}
\usepackage{dcolumn}
\usepackage{bm}
\usepackage[colorlinks=false]{hyperref} 
\usepackage{setspace}
\usepackage{amsfonts}
\usepackage{rotating}
\usepackage{makeidx}
\usepackage{amsthm}

\newcommand{\tr}{\mbox{Tr}}

\newcommand{\Diag}{\mbox{Diag}}
\newcommand{\sgn}{\mbox{sgn}}

\begin{document}

\title{Locality of three-qubit Greenberger-Horne-Zeilinger-symmetric states \footnote{Phys. Rev. A 105, 062202 (2022).}}

\author{Dian Zhu}
\affiliation{Department of Physics, School of Science, Tianjin University, Tianjin 300072, China}

\author{Gang-Gang He}
\affiliation{Department of Physics, School of Science, Tianjin University, Tianjin 300072, China}

\author{Fu-Lin Zhang}
\email[Corresponding author: ]{flzhang@tju.edu.cn}
\affiliation{Department of Physics, School of Science, Tianjin University, Tianjin 300072, China}

 \date{\today}

\begin{abstract}
The hierarchy of nonlocality and entanglement in multipartite systems is one of the fundamental problems in quantum physics.
We study this topic in three-qubit systems considering the entanglement classification of stochastic local operations and classical communication (SLOCC).
The equivalence under SLOCC divides three-qubit states into {separable},  {biseparable},  {\emph{W}}, and  {Greenberger-Horne-Zeilinger} ({GHZ})  classes.
The \emph{W} and GHZ are two subclasses of genuine tripartite entanglement.
We adopt the family of GHZ-symmetric states as a research subject,
which share the symmetries of the GHZ state and have a complete characterization of SLOCC classes.
In the biseparable region (with bipartite entanglement), there exist GHZ-symmetric states that are found to be fully local.
In addition, there are bilocal states in both the \emph{W} and GHZ classes.
That is, neither of the subclasses of genuine tripartite entanglement can ensure genuinely tripartite nonlocality.

\end{abstract}

\maketitle

 \section{Introduction}\label{Intro}

 Several concepts of nonclassical correlations in composite quantum  systems have been presented to reveal significant differences between the quantum and classical worlds   \cite{Book,RevModPhys.81.865,RMP2014bell,JPA2014LHV,RMP2012Vedral}.
 Many of them can be traced back to the early days of quantum mechanics, and play key roles in different quantum information processes.
 In general, these correlations arise from the coherent superposition of composite quantum  systems, and are equivalent to each other in pure states.
The classical probabilities in  mixed states divide them into hierarchies, and thereby lead to different classical-quantum boundaries   \cite{RMP2012Vedral}.
The hierarchy of quantum correlations is an important issue in both fundamental quantum theory and practical applications of quantum information.

Studies on the hierarchy  of quantum correlations often need carefully examining their related classicality.
 Entanglement exists in nonseparable states, while a  separable state is defined as the one that can be expressed as a mixture of product states  \cite{RevModPhys.81.865}.
Nonlocality   \cite{RMP2014bell,JPA2014LHV} is demonstrated by the absence of  local-hidden-variable (LHV) models for the outcomes of  local measurements,
which can be detected by the violation of the Bell inequality.
The nonlocal states  are a strict subset of the entangled states, as all separable states and a part of the entangled states can be modeled by LHV theories.
The existence of LHV models for separable states is  trivial, but for entangled states is an important progress, which was found by Werner   \cite{Werner1989}
in a family of bipartite mixed states, nowadays known as the Werner states.

Since Werner put forward his original results   \cite{Werner1989}, the relation between entanglement and nonlocality for bipartite quantum systems has been intensively discussed in many directions,
including  performing general measurements nonsequentially   \cite{WernerState2002} or sequentially   \cite{HiddenNonlocality} and in schemes using several copies   \cite{PRL2012MultNonlocal,PhysRevA.87.042104,PhysRevLett.100.090403,NatCommun2011,PhysRevA.72.042310}.
However, very few works  for multipartite systems were reported.
Here, the interesting but also the difficult point is that multipartite states offer more abundant structures of entanglement and nonlocality.
The notions of genuine multipartite entanglement (GME) and genuine multipartite nonlocality (GMNL) among an entire system
were proposed to distinguish them with the ones in subsystems.
Research on the locality of multipartite entangled states was initiated by T\'{o}th and Ac\'{\i}n   \cite{PhysRevA.74.030306}, who found a fully local model for a family of three-qubit states with GME.
Two recent works extended the investigation to the systems with any number of parties.
Augusiak \textit{et al.}   \cite{PhysRevLett.115.030404} showed GME and GMNL are inequivalent  by constructing a bilocal model, in which the parties are separated into two groups, for a class of GME states.
Bowles \textit{et al.}   \cite{PhysRevLett.116.130401} found that there exist states with GME admitting a fully LHV model, which can never lead to any Bell inequality violation for general nonsequential measurements.
%
%

In this work, we investigate the hierarchy of entanglement and nonlocality in three-qubit systems,
considering the classification of entanglement under stochastic local operations and classical communication (SLOCC) \cite{PhysRevA.62.062314,PhysRevA.63.012307,acin2001classification}.
Such a classification is more detailed than the scheme according  to the number of parties,
discussed in \cite{PhysRevA.74.030306,PhysRevLett.115.030404,PhysRevLett.116.130401}.
The space of three-qubit states is divided into four SLOCC classes as  {separable},  {biseparable},  {\emph{W}}, or
 {Greenberger-Horne-Zeilinger} ({GHZ})  \cite{acin2001classification}.
\emph{W} and GHZ are two inequivalent classes of genuine tripartite entanglement.
Eltschka and Siewert   \cite{eltschka2012entanglement,PhysRevLett.108.230502}
 determined exactly the SLOCC classes of a two-parameter family of three-qubit states with the same symmetry as the GHZ state,
known as GHZ-symmetric states.
The way to define these states through invariant operations is a natural  generalization of  Werner's work \cite{Werner1989} to three-qubit systems.
However,
the hierarchy of entanglement and nonlocality in such a  family of states has not been revealed by constructing LHV models as the original work \cite{Werner1989} .
%
The present work  fills this gap.
One should  expect interesting and novel phenomena due to the subtle structure of entanglement.
A natural question is here
 \emph{whether a specific type of GME, W, or GHZ, can completely rule out local models, and thus ensure GMNL}.



Our models are based on the  optimal local-hidden-state (LHS) models for Bell diagonal states  \cite{JOSAB2015Steering},
 which is an important result in the areas of Einstein-Podolsky-Rosen (EPR) steering   \cite{PRL2007Steering,RMP2020steer}
 and local models for entangled states   \cite{JPA2014quantum}.
EPR steering,
denoting a quantum correlation lying between nonlocality and entanglement,
exists in the entangled states whose unnormalized postmeasured states, after one-sided local measurements, cannot be described by a LHS model \cite{PRL2007Steering,RMP2020steer}.
The EPR steering is not only an important resource in quantum information processes   \cite{RMP2020steer,PRA2012OneSDIQKD,JPA2014quantum},
but also acts as a powerful tool in research of entanglement and nonlocality,
such as in the construction of counterexamples to the Peres conjecture   \cite{Peres1999,NP2014,PhysRevLett.113.050404},
generalizing Gisin's theorem   \cite{Gisin1991,SR2015chen} and designing the algorithms for LHV models   \cite{Arxiv2015LHV, PRL2016Algorithmic}.
The present work offers an alternative example for this point.

%


%

The technique to construct local models in this work is divided into two steps.
First, by replacing the hidden state in the mentioned optimal model with a two-qubit state, we construct a local model for a tripartite state.
This is feasible as a LHS model is a particular case of a LHV model with the hidden variable being a local state.
Such a replacement was used in the work of Augusiak \textit{et al.}   \cite{PhysRevLett.115.030404}.
Generally speaking, it is difficult to determine the entanglement class of the tripartite state.
Second, we perform the GHZ symmetrization   \cite{eltschka2012entanglement,PhysRevLett.108.230502} on the tripartite state and its local model simultaneously,
and thereby obtain a GHZ-symmetric state admitting a LHV model.
Our results show that, a biseparable GHZ-symmetric state with two-qubit entanglement may be fully local,
and there exist bilocal states in both the \emph{W} and GHZ classes.
The latter offers a negative answer to the  question above.

\section{Two-qubit GHZ-symmetric states}\label{qubit}

Let us begin with the two-qubit GHZ-symmetric states introduced by Siewert and Eltschka   \cite{PhysRevLett.108.230502}.
Although the LHV models we present in this part are trivially reformulated from the the existing LHS models,
the following results will serve as important tools for the tripartite case.

\begin{figure}
 \centering
 \includegraphics[width=7.5cm]{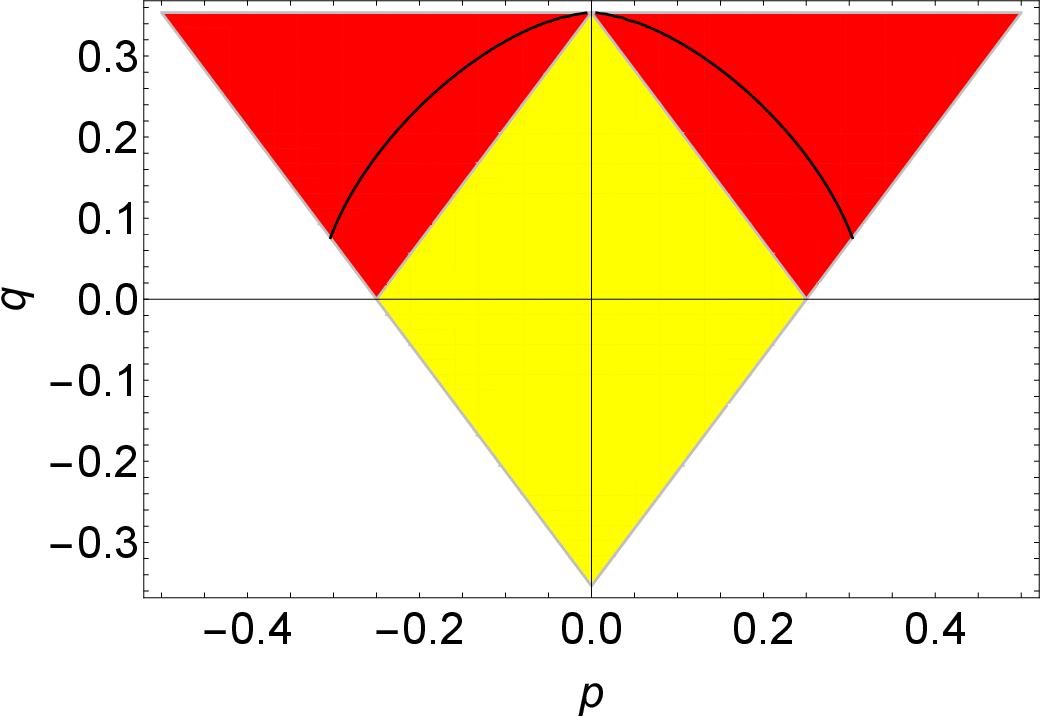}
 \caption{
Triangle of  two-qubit GHZ-symmetric states, constrained by  $\frac{{q}}{\sqrt{2}} \!\pm\! {p}\!+\!\frac{1}{4} \!\ge\! 0$ and $-\frac{1}{2\sqrt{2}}\! \leq \!{q}\! \leq \!\frac{1}{2\sqrt{2}}$ .
 The upper corners of the triangle are the Bell states  $|\phi^{+}\rangle$ and $|\phi^{-}\rangle$.
 The maximally mixed state $\frac{\openone}{4}$  locates at the origin.
The entangled (red) and separable (yellow) regions are divided by the lines of $q=\frac{1}{2\sqrt{2}}\pm\sqrt{2}p$.
 The black solid curve 
 corresponds to the LHV model in Eq. (\ref{lhvPab}) and satisfies the function (\ref{xy}).
}\label{twoqlhvmodel}
\end{figure}

The GHZ-symmetric states of a two-qubit system share the symmetries of the two Bell states $\left|\phi^{\pm}\right\rangle\!=\!\frac{1}{\sqrt{2}}(|00\rangle \pm |11\rangle)$ and can be written as
\begin{equation}  \label{rhoS}
\begin{split}
\rho^S&=(\sqrt{2}q+p)\left|\phi^+\right\rangle\left\langle\phi^+\right|+(\sqrt{2}q-p)\left|\phi^-\right\rangle\left\langle\phi^-\right|\\
&+(1-2\sqrt{2}q) \frac{\openone}{4},
\end{split}
\end{equation}
 where the two parameters    $-\frac{1}{2\sqrt{2}}\! \leq \!{q}\! \leq \!\frac{1}{2\sqrt{2}}$ and  $\frac{{q}}{\sqrt{2}} \!\pm\! {p}\!+\!\frac{1}{4} \!\ge\! 0$ .
They form a convex set that defines a triangle on the plane of $(p,q)$ as shown in Fig. \ref{twoqlhvmodel}.
 The separable boundary are two lines of $q=\frac{1}{2\sqrt{2}}\pm\sqrt{2}p$.
The triangle is actually equivalent to a cross section of the tetrahedron for Bell diagonal states
in the space of $(T_x,T_y,T_z)$   \cite{horodecki1996information,dakic2010necessary}.
Here,  the Bell diagonal states are of the form
\begin{equation}
\rho^{B}\!=\!\frac{1}{4}\left[\openone+({T}\,\vec{\sigma})\cdot \vec{\sigma}\right],
\end{equation}
whose spin correlation matrix is diagonal as ${T}\!=\!\Diag [ T_{x},T_{y},T_{z} ]$,
with $\vec{\sigma}=(\sigma_x$, $\sigma_y$, $\sigma_z)^{\mathrm{T}}$ being the vector of Pauli operators.
The matrices of GHZ-symmetric states (\ref{rhoS}) satisfy $T_{x}\!=\!-T_{y}\!=\!2{p}$ and $T_{z}\!=\!2\sqrt{2}{q}$.

The optimal LHS models for Bell diagonal states were originally constructed by Jevtic \textit{et al.}   \cite{JOSAB2015Steering} based on the steering ellipsoid \cite{PRL2014Ellips}.
These models drew an EPR-steerable boundary in the space of $(T_x,T_y,T_z)$.
Zhang and Zhang   \cite{ZhangPRA2019} showed a simple approach to generate them from Werner's results.
We here follow the formulas in the later work.

Let $\rho $ denote a  state  of the whole system of $A$ and $B$  and $ \Pi^{\vec{x}}_a  $ be a projector of a measurement labeled by $\vec{x}$ corresponding to outcome $a$.
After a local measurement on subsystem $A$, the unnormalized postmeasured state of $B$   is
$
\rho_{a}^{\vec{x}}=\tr_A  (\Pi^{\vec{x}}_a \otimes \openone \rho),
$
where
$\tr_A$ is the partial trace over $A$.
A LHS model is defined as
\begin{equation} \label{LHS}
\rho^{\mathrm{LHS}}=\int\omega(\vec{\lambda})P_{A}(a|\vec{x},\vec{\lambda})\rho_{\lambda}\,d\vec{\lambda},
\end{equation}
where $\vec{\lambda}$ represents a hidden variable with a distribution $\omega(\vec{\lambda})$,
 $\rho_{\lambda}$ is a hidden state depending on $\vec{\lambda}$, and $P_{A}(a|\vec{x},\vec{\lambda})$ is the probability of outcome $a$ under the condition of $\vec{x}$ and $\vec{\lambda}$.
If there exists a LHS model satisfying
$
\rho_{a}^{\vec{x}}= \rho^{\mathrm{LHS}},
$
for all the measurements, the state $\rho$ is unsteerable from $A$ to $B$.

Suppose that $T_0$ is the spin correlation matrix of a Bell diagonal state on the EPR-steerable boundary.
The projector for a qubit can be expressed as $\Pi_{a}^{\vec{x}}=\frac{1}{2}(\openone+a\,\vec{x}\cdot\vec{\sigma})$ with $\vec{x}$ being a unit vector and $a=\pm1$.
The corresponding postmeasured states can be derived as
\begin{eqnarray}
\rho_{a}^{\vec{x}}=\dfrac{1}{4}\left[\openone+a (T_{0}\vec{x}) \cdot \vec{\sigma}\right].
\end{eqnarray}
In the optimal LHS models, the hidden variable is a vector $\vec{\lambda}$ on a unit sphere.
Its distribution, the conditional probability, and the corresponding hidden state are given by
\begin{eqnarray}\label{LHS2}
&& \omega(\vec{\lambda})=\frac{|T_{0}\vec{\lambda}|}{2\pi},   \nonumber \\
&&P_{A}(a|\vec{x},\vec{\lambda})=\frac{1}{2}[1+a\,\sgn(\vec{x}\cdot\vec{\lambda})],    \\
&& \rho_{\lambda}=\frac{1}{2}(\openone+\vec{\lambda}'\cdot\vec{\sigma}),  \nonumber
\end{eqnarray}
with $\vec{\lambda}' = T_0\vec{\lambda}/|T_0\vec{\lambda}|$.
Substituting them into the integral (\ref{LHS}), one can find that they satisfy
$
\rho_{a}^{\vec{x}}= \rho^{\mathrm{LHS}},
$
when the normalization condition
\begin{equation}\label{normalization}
\int\!  {|T_{0}\vec{\lambda}|} d\vec{\lambda}={2\pi},
\end{equation}
is fulfilled.
In addition, the EPR-steerable boundary is determined by the normalization condition (\ref{normalization}).
One can trivially construct a LHV model here by introducing the response function for part $B$
\begin{equation}\label{lhsPb}
P_{B}(b|\vec{y},\vec{\lambda})=\tr(\Pi_{b}^{\vec{y}} \rho_{\lambda}),
\end{equation}
where the projection operator  $\Pi_{b}^{\vec{y}}$ has the similar definition as  $\Pi_{a}^{\vec{x}}$.
Then, the outcomes of local measurements
 can be simulated by the  LHV model as
\begin{equation} \label{lhvPab}
\begin{split}
P(a , b|\vec{x} , \vec{y}) &=\tr(\Pi_{a}^{\vec{x}} \otimes \Pi_{b}^{\vec{y}} \rho^B) \\
& = \int\!\omega(\vec{\lambda})P_{A}(a|\vec{x},\vec{\lambda})P_{B}(b|\vec{y},\vec{\lambda})\,d\vec{\lambda} .
\end{split}
\end{equation}

We can explicitly solve the integral for the GHZ-symmetric states (\ref{rhoS}) and express it as
\begin{equation}\label{xy}
\frac{1}{|2{p}|}= |w| +\dfrac{\mathrm{arccosh} |w| }{\sqrt{w^2-1}}  ,
\end{equation}
with $w= {\sqrt{2}{q}}/{{p}}$.
It is the EPR-steerable boundary on the plane of $(p,q)$ displayed as the black curve in Fig. \ref{twoqlhvmodel}.
The states
between the curve and the separable boundary can be sufficiently determined to be entangled but without nonlocality.

\section{Three-qubit GHZ-symmetric states}

\begin{figure}
 \centering
 \includegraphics[width=7.5cm]{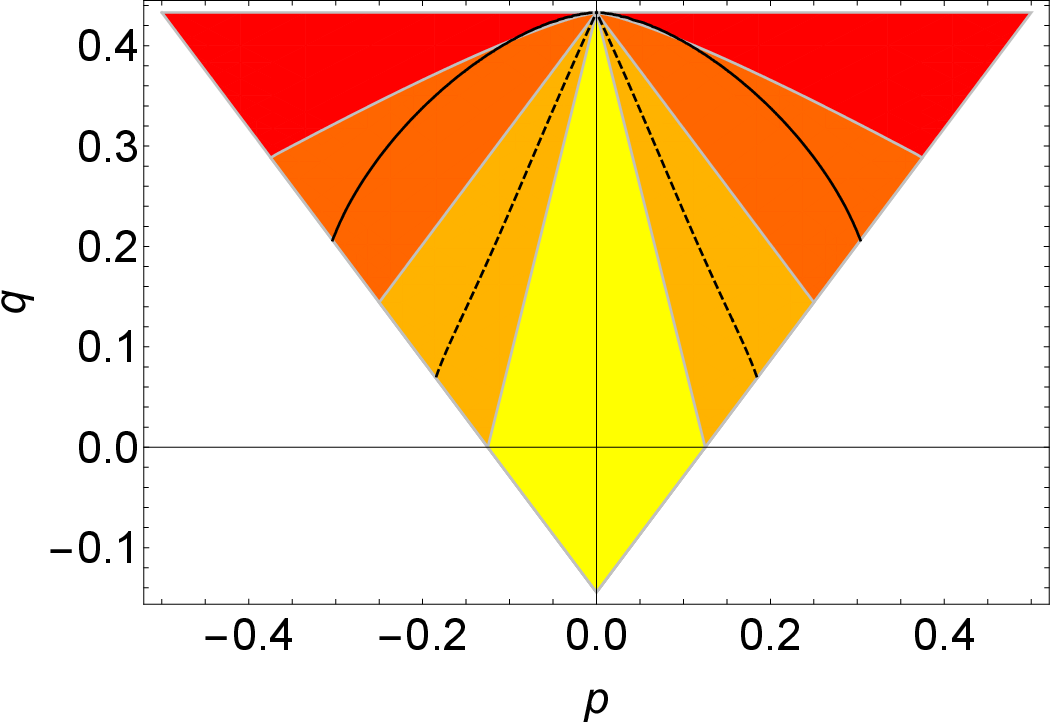}
 \caption{
Triangle of tripartite GHZ-symmetric states.
The standard GHZ states $|G_{\pm}\rangle$ are located at the two upper corners, and the maximally mixed state $\frac{\openone}{8}$ at the origin.
Different classes of entanglement (separable, biseparable, \emph{W} and GHZ) are indicated from yellow to red in order.
Their boundaries in order  are defined by: $|p|\!=\!-\frac{\sqrt{3}}{6}q+\frac{1}{8} $ (separable/biseparable), $ |p|\!=\!-\frac{\sqrt{3}}{2}q+\frac{3}{8} $ (biseparable/\emph{W}), and the parametric  curve with $p\!=\!(v^5+8v^3)/[8(4-v^2)]$ and $q\!=\![\sqrt{3}(4-v^2-v^4)]/[4(4-v^2)]$ (\emph{W}/GHZ).
The black solid curve corresponds to $\rho^S [\rho_1 (T_0)]$, which admits a bilocal model, and satisfies Eq. (\ref{xy}) but with $w=  ({2\sqrt{3}q-1/2})/{|2p|}$.
The dashed curve corresponds to $\rho^S [\rho_2 (T_1)]$, which admits a fully LHV model, and satisfies the  parametric equation (\ref{paraeq}).
 }\label{GHZ3}
\end{figure}

We now turn to the three-qubit GHZ-symmetric states.
These states share the following symmetries of the two GHZ states $\left|G_{\pm}\right\rangle\!=\!\frac{1}{\sqrt{2}}(|000\rangle \pm |111\rangle)$:
 (i) qubit permutations;
 (ii) simultaneous three-qubit flips (i.e., the application of $\sigma_{x} \otimes\sigma_{x}\otimes\sigma_{x}$); and
 (iii) qubit rotations about the $z$ axis of the form $U\left(\phi_{1},\phi_{2}\right)=e^{i\phi_{1}\sigma_{z}}\otimes e^{i\phi_{2}\sigma_{z}}\otimes e^{-i(\phi_{1}+\phi_{2})\sigma_{z}}$.
They are of the form
\begin{equation}\label{RhoS}
\begin{split}
\rho^S  =& \left( \! \frac{2 q }{\sqrt{3}}\!+\! p \! \right)\left|  G_{+}   \right\rangle\left\langle   G_{+}  \right|+ \left( \! \frac{2 q }{\sqrt{3}}\!-\! p \! \right)\left|   G_{-}   \right\rangle\left\langle    G_{-}  \right| \\
&+\left(\! 1 \!- \!\frac{4 q }{\sqrt{3}}\! \right)   \frac{\openone}{8},
\end{split}
\end{equation}
where the two parameters satisfy $-\frac{1}{4\sqrt{3}}\leq q \leq \frac{\sqrt{3}}{4}$ and
$
|p| \leq \frac{1}{8}+\frac{\sqrt{3}}{2} q
$.
As shown in Fig. \ref{GHZ3}, this family of states forms an isosceles triangle in the state space.
Two pure states $\left|G_{\pm}\right\rangle $ are located at the two upper corners and the origin corresponds to the maximally mixed state $\frac{\openone}{8}$.
Eltschka and Siewert   \cite{eltschka2012entanglement} divided the triangle into four regions according to the entanglement classes as fully separable, biseparable, \emph{W}, or GHZ,
which are displayed by different colors in Fig. \ref{GHZ3}.

For an arbitrary three-qubit state $\rho$,
one can derive a corresponding GHZ-symmetric state by performing a GHZ symmetrization
\begin{equation}\label{GHZsz}
\rho^S(\rho)=\int\,du(u\rho u^\dagger),
\end{equation}
where the integral is  to cover the entire GHZ symmetry group, i.e., the operations in (i), (ii), (iii),  and their products.
The coordinates of $\rho^S(\rho)$ can be inferred from two matrix elements of $\rho$ as
\begin{equation}\label{trippq}
\begin{split}
p&=\frac{1}{2}\biggr(\!\left\langle {G}_{+}\right|\!\rho \!\left| {G}_{+}\right\rangle\!-\!\left\langle {G}_{-}\right|\!\rho \!\left| {G}_{-}\right\rangle\!\biggr),\\
q&= \frac{1}{\sqrt{3}}\biggr(\!\left\langle {G}_{+}\right|\!\rho \!\left| {G}_{+}\right\rangle\!+\!\left\langle {G}_{-}\right|\!\rho \!\left| {G}_{-}\right\rangle\!-\!\frac{1}{4}\!\biggr).
\end{split}
\end{equation}
 We remark that, to obtain a state $\rho^S(\rho)$, the \textit{state} $\rho$ does not need to be a state.
 Namely, the operator $\rho$  is required to be Hermitian and normalized,
 but the positive semi-definitiveness can be replaced with the  condition that its corresponding $(p,q)$ are in the physical region.
In the following parts, we actually utilize such \textit{states} to construct local GHZ-symmetric states.

The nonlocality (GMNL, bilocality, or full locality) of $\rho$ is preserved by the operations in (i), (ii), and (iii), which are local unitary or permutations.
Consequently, the GHZ symmetrization only could reduce or preserve the nonlocality.
Here, to reduce means to change the nonlocality into a weaker one, e.g., to change GMNL into a bilocality or full locality.
Said in a different way, if the \textit{state} $\rho$ admits a LHV model, by symmetrizing the model, we can construct one for $\rho^S(\rho)$.
Based on this idea and the results in the two-qubit case, we give the following models for $\rho_S$ and sufficient criterions for its locality.

 \subsection{Bilocal model}\label{BLHV}

A direct result can be obtained by replacing the basis $\{|0\rangle$,$|1\rangle\}$ of $B$ in two-qubit GHZ symmetric states with $\{|00\rangle$, $|11\rangle\}$ of systems  $B$ and  $C$.
This leads to a family of three-qubit \textit{states} for $ABC$
\begin{equation}\label{rhothree}
\rho_1(T)=\frac{1}{4}\left[  \openone \otimes\Sigma_{0}+\left (T\vec{\sigma} \right)\cdot \vec{\Sigma}\right],
\end{equation}
where $\Sigma_{0} =|00\rangle\langle00|+|11\rangle\langle11|, \Sigma_{x} =|00\rangle\langle11|+|11\rangle\langle00|, \Sigma_{y} =-i|00\rangle\langle11|+i|11\rangle\langle00|$, and $\Sigma_{z} =|00\rangle\langle00|-|11\rangle\langle11|$ are  the generalized Pauli matrices in the subspace of  $\{|00\rangle$, $|11\rangle\}$.
When $T=T_0$ satisfy the normalization condition (\ref{normalization}) and thereby on the EPR-steerable boundary,
The tripartite states naturally admit a family of LHV models as
\begin{equation} \label{lhvPabc}
\begin{split}
\! \! \! P_b(a , \! b, \!c|\vec{x} , \vec{y},\vec{z}) & \!= \! \tr\left[\Pi_{a}^{\vec{x}} \otimes \Pi_{b}^{\vec{y}} \otimes \Pi_{c}^{\vec{z}} \rho_1(T_0) \right] \\
& \!=\! \int \! \! \!\omega(\vec{\lambda})P_{\!A}(a |\!\vec{x},\!\vec{\lambda})P_{\!BC}(b,\!c | \vec{y},\!\vec{z},\!\vec{\lambda})\,d\vec{\lambda},
\end{split}
\end{equation}
where $P_{BC}(b,c|\vec{y},\vec{z},\vec{\lambda})   = \tr(\Pi_{b}^{\vec{y}}  \otimes \Pi_{c}^{\vec{z}} \rho_{\lambda}^{BC} ) $, $\rho_{\lambda}^{BC}=\frac{1}{2}(\Sigma_{0}\!+\!\vec{\lambda}'\!\cdot\!\vec{\Sigma})$, and other parameters and the functions are defined in Eq. (\ref{LHS2}).
Since the bipartite state $\rho_{\lambda}^{BC} $ is an entangled pure state in general, the response function $P_{BC}(b,c|\vec{y},\vec{z},\vec{\lambda}) $ can not be further decomposed.
The models are bilocal, which exclude GMNL in the states  but allow bipartite nonlocality between $B$ and  $C$.

One can now symmetrize the states $\rho_1(T_0)$  into the GHZ-symmetric family.
Actually, only the permutations are necessary, as the states are invariant under three-qubit flips and the rotations $U(\phi_1,\phi_2)$.
The corresponding GHZ-symmetric states $\rho^S[\rho_1(T_0)]$ draw a curve, which can also be parameterized as in Eq. (\ref{xy}) but with $w=  ({\sqrt{3}q-1/4})/{p}$, on the plane of $(p,q)$ as shown in Fig. \ref{GHZ3}.
They admit a bilocal model which can be expressed as an average of  the one in Eq. (\ref{lhvPabc}) and its two permutations of ${A\leftrightarrow B}$ and ${A\leftrightarrow C}$. We omit the formula here.

The convex combinations of $\rho^S[\rho_1(T_0)]$ and fully separable states form the region under the curve for  $\rho^S[\rho_1(T_0)]$  in the triangle,
in which the GMNL is certainly excluded.
It is worth noting that the curve passes through both \emph{W} and GHZ entanglement,
 although it is just marginally higher than the \emph{W}/GHZ  boundary  near the state $\rho^S=\frac{1}{2}(|G_+\rangle\langle G_+|+ |G_-\rangle\langle G_-|)$.
Consequently,  both the two classes of  genuine three-qubit entanglement, \emph{W} and GHZ, are inequivalent to  genuine three-qubit nonlocality.

\begin{figure}
 \centering
 \includegraphics[width=7.5cm]{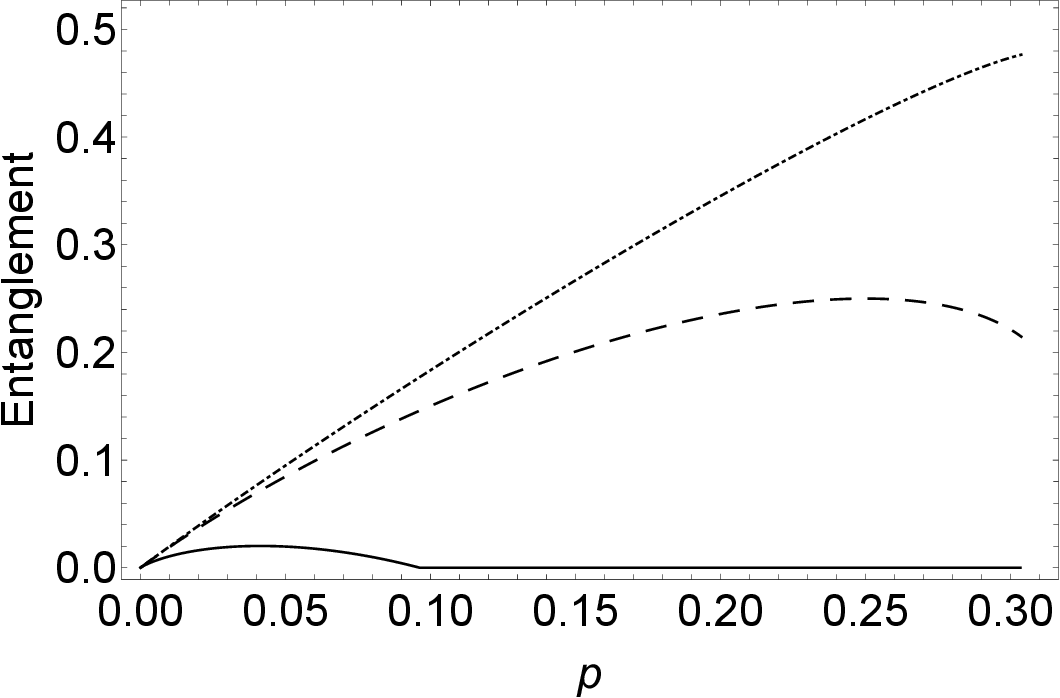}
  \caption{
Entanglement degrees vs.  the $p$ coordinate of  $\rho^S[\rho_1(T_0)]$ .
The curves show the amounts of  total entanglement, GME, and GHZ entanglement in order from top to bottom.
 }\label{EntBilocal}
\end{figure}

One can observe the deviations of $\rho^S[\rho_1(T_0)]$ from the entanglement boundaries by deriving its entanglement degrees, shown in Fig. \ref{EntBilocal}.
We adopt three measures for the entanglement in GHZ-symmetric states.
The three-tangle is a natural choice to quantify the GHZ entanglement, as Siewert and Eltschka   \cite{PhysRevLett.108.230502} derived its analytical results for the GHZ-symmetric states.
Given a state with coordinates $(p,q)$ ($p\geq0$, without loss of generality), its three-tangle is given by
\begin{equation}\label{tangle}
\tau_{3}(p,q)= \max \left\{0, \frac{p-p^{W}}{\frac{1}{2}-p^{W}} \right\},
\end{equation}
where $p^{W}$ is the $p$ coordinate of the intersection point of the straight line, connecting $(p,q)$ and the corner for $|G_+\rangle$, and the \emph{W}/GHZ  boundary.
The GHZ-symmetric states belong to the family of  three-qubit $X$ matrices.
Their total entanglement (opposite to full separability) and GME can be measured by two generalizations of concurrence   \cite{PHDGME,GMC}, which can be expressed
by using the coordinates as
 \begin{equation}
 \begin{split}
 &\mathrm{C}_{T}=\max\left\{0,2|p|+\frac{\sqrt{3}}{3}q-\frac{1}{4}\right\},\\
 &\mathrm{C}_{G}=\max\left\{0,2|p|+\sqrt{3}q-\frac{3}{4}\right\}.
\end{split}
 \end{equation}
Similar to the  three-tangle (\ref{tangle}), both of them can be geometrically interpreted as a relative distance to the corresponding boundary.
As shown in Fig. \ref{EntBilocal}, the GHZ entanglement has a nonzero section near its left endpoint.
The other two degrees of entanglement are always greater than zero,
and total entanglement monotonously increases with  $\rho^S[\rho_1(T_0)]$ moving from the top line of this triangle to the right side.

\subsection{Fully local model}

Our approach can be extended to construct a fully LHV model, by modifying the hidden state $\rho_{\lambda}^{BC}$ to a local one.
That is, when $\rho_{\lambda}^{BC}$ admits a local model as
\begin{equation} \label{lhvPbc}
\begin{split}
\!\!\! \!\!\!  P_{BC}(b , c|\vec{y} , \vec{z},\vec{\lambda}) &\!  =\!\tr(\Pi_{b}^{\vec{y}} \otimes \Pi_{c}^{\vec{z}}\rho_{\lambda}^{BC}) \\
&\!=\!  \int\!\!\tau(\vec{\mu},\!\vec{\lambda})P_{\!B}(b| \vec{y}, \vec{\mu}, \vec{\lambda})P_{\!C}(c| \vec{z}, \vec{\mu}, \vec{\lambda})\,d\vec{\mu}, \nonumber
\end{split}
\end{equation}
with a hidden variable $\vec{\mu}$ and a distribution $\tau(\vec{\mu}, \vec{\lambda})$, the integral in Eq. (\ref{lhvPabc}) presents a fully LHV model as
\begin{equation} \label{flhv}
\begin{split}
\! \! \! P_f(a , \! b, \!c|\vec{x} , \vec{y},\vec{z})
  \!=\! \int \! \! \!\omega(\vec{\lambda})P_{\!A}(a |\!\vec{x},\!\vec{\lambda})P_{\!BC}(b,\!c | \vec{y},\!\vec{z},\!\vec{\lambda})\,d\vec{\lambda}.
\end{split}
\end{equation}
Here, the hidden variable is $(\vec{\lambda}, \vec{\mu})$ in a joint distribution $\Omega(\vec{\lambda}, \vec{\mu})=\omega(\vec{\lambda})\tau(\vec{\mu},\vec{\lambda})$.

To  analytically construct a fully local model for GHZ-symmetric states,
we extend the three-qubit \textit{states} (\ref{rhothree}) to the form
\begin{equation}
\rho_2(T)\!=\!\dfrac{1}{4}\left[t \openone \otimes\Sigma_{0}\!+\!(T\vec{\sigma})\cdot  \vec{\Sigma}    \right]\! +\frac{1}{2}(1-t) \openone \otimes D_{01},
\end{equation}
where $t\in[0,1]$ and $ D_{01}$ is a normalized diagonal state in the subspace of $\{ |01\rangle , |10\rangle\}$.
The coordinates of $\rho^S[\rho_2(T)]$ are determined by the first part of $\rho_2(T)$ as
\begin{equation}
p=\frac{1}{2}T_x, \ \ q=\frac{1}{2\sqrt{3}}\left(t+T_z-\frac{1}{2}\right).
\end{equation}

We now present a one-parameter family of fully LHV models corresponding to a family of states in the above form.
Suppose that the parameter and matrix  in the states are $t=t_1$ and $T=T_1$.
The hidden variable is still chosen as a vector $\vec{\lambda}$ on a unit sphere.
 We define its distribution, the conditional probability, and the corresponding hidden state as
\begin{eqnarray}
&& \omega(\vec{\lambda})=\frac{|T_{1}\vec{\lambda}|}{2\pi} \left[1+ q (\vec{\lambda} )\right] ,   \nonumber \\
&&P_{A}(a|\vec{x},\vec{\lambda})=\frac{1}{2}\left[1+a\,\sgn (\vec{x}\cdot\vec{\lambda} )\right],    \\
&& \rho_{\lambda}^{BC}=\frac{1}{1+ q(\vec{\lambda})}\left[\frac{1}{2}\left(\Sigma_{0}+\vec{\lambda''}\cdot\vec{\Sigma}\right)+ q (\vec{\lambda} ) D_{01}\right],  \nonumber
\end{eqnarray}
where $\vec{\lambda''}=  T_{1}\vec{\lambda} /|T_{1}\vec{\lambda}|$, $q (\vec{\lambda} )  \in[0,1]$ and $q (-\vec{\lambda} ) =q (\vec{\lambda} ) $.
Substituting them into Eq. (\ref{flhv}) and requiring
\begin{equation}
  P_f(a , \! b, \!c|\vec{x} , \vec{y},\vec{z})= \tr\left[\Pi_{a}^{\vec{x}} \otimes \Pi_{b}^{\vec{y}} \otimes \Pi_{c}^{\vec{z}} \rho_2(T_1)\right] ,
\end{equation}
one can obtain the relations
\begin{equation} \label{t1T1}
  \int\!\frac{|T_{1}\vec{\lambda}|}{2\pi} \,d\vec{\lambda} = t_1, \ \
 \int \!\frac{|T_{1}\vec{\lambda}|}{2\pi} q(\vec{\lambda}) \,d\vec{\lambda} =1-t_1.
\end{equation}
Therefore, a given form of $q(\vec{\lambda}) $ can determine a curve on the triangle.
We choose $q(\vec{\lambda}) = \min\{ \sin\theta'', c=1-w_c \}$, where $\theta''$ is the angle between the positive \emph{z} axis and $\vec{\lambda''}$,
 and  $w_c \approx 0.354$ is the value of $w$ on the intersection point of the curve (\ref{xy}) and the side of the triangle for the two-qubit case.
 The definition of $D_{01}$ corresponding to the two options is $D_{01} =\frac{1}{2}(|01\rangle \langle 01| + |10\rangle \langle 10|) $  for $q(\vec{\lambda}) =  \sin\theta''$ and $D_{01} = \cos^2 \frac{\theta''}{2}|01\rangle \langle 01| + \sin^2 \frac{\theta''}{2}|10\rangle \langle 10| $  for $q(\vec{\lambda}) =  1-w_c$.
 In the first case, $\rho_{\lambda}^{BC}$ is separable.
In the second case, $\rho_{\lambda}^{BC}$ is unsteerable, as it is equivalent to  the  state on the above-mentioned intersection point, under the local positive linear map  $\mathcal{M}=(e^{i \phi''}\sin \frac{\theta''}{2}|0\rangle \langle 0| + \cos\frac{\theta''}{2}|1\rangle \langle 1|) \otimes \openone$ \cite{arxiv2015UnSteer}.

 The two integrals in Eq. (\ref{t1T1}) can be explicitly solved, even though $q(\vec{\lambda})$ is a piecewise function.
Direct calculation gives the coordinates of $\rho^S[\rho(T_1)]$ as a parametric equation
\begin{equation}\label{paraeq}
p=\frac{1}{2 G(v)} , \ \ q=\frac{1}{2\sqrt{3}}\left[\frac{G_0(v) + v}{G(v)}-\frac{1}{2}\right],
\end{equation}
where $G(v)=G_0(v)+G_1(v)$, and $G_0(v)$ and $G_1(v)$ are
 two functions of $v$,
 given by
\begin{equation}
\begin{split}
G_{0}(v)&\!=\!\frac{\mathrm{arcsinh} \sqrt{v^2-1}}{\sqrt{v^2-1}} + \sqrt{v^2},\\
G_{1}(v)&\!=\! c \frac{\mathrm{arcsinh} \sqrt{\frac{(1-c^2)(v^2-1)}{1-c^2(1-v^2)}}}{\sqrt{v^2-1}}\!+\!\arctan \sqrt{\frac{c^2 v^2}{1-c^2}}.
\end{split}
\end{equation}
The one-parameter family of states $\rho^S[\rho(T_1)]$ draws a curve on the triangle,
the region under which are of  fully local   states.
This shows that the bipartite entanglement is not equivalent to bipartite nonlocality in the GHZ-symmetric states.

\section{Four-qubit restricted GHZ-symmetric states}

\begin{figure}
 \centering
  \includegraphics[width=7.5cm]{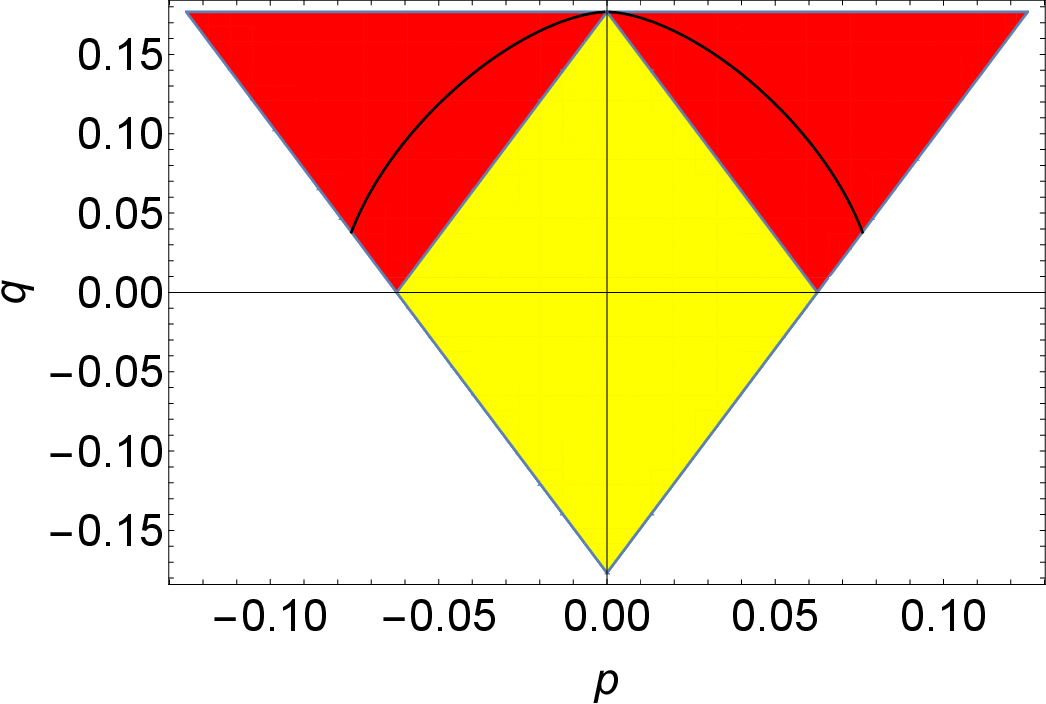}
  \caption{
The region of four-qubit RGHZ symmetric states forms a  triangle with  $q \geq \pm 2\sqrt{2}p - \sqrt{2}/8$ and $|p| \leq 1/8$.
  The maximally mixed state $\frac{\openone}{16}$ is at the origin.
  The $G_{abcd}$ class (red) and the $L_{abc_{2}}$ class (yellow) are separated by the line $q=\pm 2\sqrt{2}p +\sqrt{2}/8$.
The black solid curve satisfies the parametric equation (\ref{xy}) but with $ \sqrt{2}w=q /p$.
  }   \label{RGHZfig}
\end{figure}

Our approach to construct local models can be directly applied in highly symmetric states with more qubits.
Let us take a family of bilocal models for
four-qubit restricted GHZ-symmetric (RGHZ-symmetric) states  \cite{PhysRevA.89.052326}, for example.
The RGHZ-symmetric states are a two-parameter family of four-qubit states,
which are invariant under simultaneous two-qubit flips and four-qubit GHZ symmetric operations.
They involve only two of the nine degenerate SLOCC entanglement classes for four-qubit states,
 denoted by $G_{abcd}$ and $L_{abc_{2}}$  \cite{PhysRevA.89.052326,PRA.65.052112}.
The general form of RGHZ-symmetric states is
\begin{equation}
\begin{split}
\rho^{RS} = &p ( |0000 \rangle\langle 1111| \! + \! |1111 \rangle\langle 0000| )  \\
&+ \Diag (\alpha_1,\alpha_2,\alpha_2,\alpha_1 , \alpha_2,\alpha_1,\alpha_1,\alpha_2 , \\
& \ \ \ \ \ \ \ \ \ \ \ \ \alpha_2,\alpha_1,\alpha_1,\alpha_2 , \alpha_1,\alpha_2,\alpha_2, \alpha_1)
\end{split}
\end{equation}
with   $\alpha_1 = \frac{1}{16} + \frac{q}{2\sqrt{2}}$, $\alpha_2 = \frac{1}{16} - \frac{q}{2\sqrt{2}}$.
Figure \ref{RGHZfig} shows the region of RGHZ-symmetric states  on the plane of $(p,q)$,
 with the  $G_{abcd}$ and $L_{abc_{2}}$ classes displayed by different colors.
By replacing the basis $\{|0\rangle , |1\rangle\}$ of $B$ in two-qubit GHZ symmetric states on the EPR-steerable boundary
with $\{|000\rangle , |111\rangle\}$ of systems $B$, $C$, and $D$,
one obtains a family of four-qubit states as
\begin{equation}
\quad \rho_3 (T_0) = \frac{1}{4}(\openone \otimes \Sigma_0 '+ T_0 \vec{\sigma}\cdot\vec{\Sigma'}).
\end{equation}
Here $\Sigma_i '$ ($i=0, x, y, z$)  are the generalized Pauli matrices in the subspace of $\{|000\rangle,|111\rangle\}$, similar to the ones in Sec. \ref{BLHV}.
Such states naturally admit a family of LHV model as
\begin{equation}\label{LHV4}
\begin{split}
&   \mathrm{Tr}  \left[  \Pi_{a}^{\vec{x}}   \otimes   \Pi_{b}^{\vec{y}}   \otimes   \Pi_{c}^{\vec{z}}   \otimes   \Pi_{d}^{\vec{t}}   \rho_3   (T_0)   \right]   \\
&=   \int   \omega  (  \vec{\lambda}  ) P_A   ( a | \vec{x}  , \vec{\lambda} ) \mathrm{Tr}\left(\Pi_{b}^{\vec{y}} \otimes \Pi_{c}^{\vec{z}} \otimes \Pi_{d}^{\vec{t}}
\rho_{\lambda}^{BCD}\right) d\vec{\lambda},
\end{split}
\end{equation}
where $\vec{\lambda} $, $ \omega  (  \vec{\lambda}  )$ and $ P_A   ( a | \vec{x}  , \vec{\lambda} ) $
are defined in Eq. (\ref{LHS2}), and
$\rho_{\lambda}^{BCD} = \frac{1}{2} (\Sigma_0 + \vec{\lambda} ' \cdot \vec{\Sigma})$
with $\vec{\lambda}' = T_0\vec{\lambda}/|T_0\vec{\lambda}|$.
%
%
%
One can perform a  RGHZ symmetrization on  $\rho_3 (T_0)$ and the model (\ref{LHV4}) simultaneously,
and obtains a one-parameter RGHZ-symmetric state with a bilocal model,  shown by the  curve in Fig. \ref{RGHZfig}.
The bilocal  model is in the form of \emph{1 versus 3} as
\begin{equation}
\begin{split}
& \ \ \int \! \!     \omega_1  (  \vec{\lambda} _1  ) P_A   ( a | \vec{x}  , \vec{\lambda} _1  ) P_{BCD}   ( b,c,d | \vec{y},\vec{z} ,\vec{t}   , \vec{\lambda} _1  ) d\vec{\lambda}_1\\
& \!  + \!  \int   \! \!    \omega_2  (  \vec{\lambda} _2  ) P_B   ( b | \vec{y}  , \vec{\lambda} _2  ) P_{CDA}   ( b,c,a | \vec{z} ,\vec{t} ,\vec{x}  , \vec{\lambda} _2 ) d\vec{\lambda}_2\\
& \!  +\!  \int   \! \!     \omega_3  (  \vec{\lambda} _3  ) P_C   ( c | \vec{z}  , \vec{\lambda} _3  ) P_{DAB}   (d,a,b | \vec{t} ,\vec{x} ,\vec{y}   , \vec{\lambda} _3  ) d\vec{\lambda}_3\\
&\!   +\!  \int   \! \!     \omega_4 (  \vec{\lambda} _4  ) P_D   ( d | \vec{t}  , \vec{\lambda} _4  ) P_{ABC}   (a,b,c | \vec{x}   ,\vec{y},\vec{z}    , \vec{\lambda} _4  ) d\vec{\lambda}_4,
\end{split}
\end{equation}
while \emph{there is no one-qubit tensor product three-qubit entangled states in the RGHZ-symmetric states }  \cite{PhysRevA.89.052326}.
This fact suggests a difference between biseparablity and bilocality in four-qubit states.


%


\section{Summary}\label{summ}

We investigate the relations between entanglement and nonlocality in multipartite systems considering the SLOCC classification of entanglement.
However, both determining entanglement properties and constructing local models for a given set of  multipartite states are two challenges in theoretic studies.
We adopt a two-parameter family of three-qubit states, the so-called GHZ-symmetric states,  as a research object,
as the entanglement of these states have an exact description.
By performing the GHZ symmetrization on the tripartite extensions of two-qubit local model, we present two one-parameter families of LHV models.
These models show that both the bipartite entanglement and genuine tripartite entanglement are inequivalent to the nonlocality at the same level.
In particular, there exist bilocal states in both the two subclasses, \emph{W} and GHZ, of genuinely tripartite entangled states,
which do not display genuine tripartite nonlocality.


It would be interesting to extend our results in the following two directions.
On the one hand,
our approach to construct LHV models can be directly applied to the systems with more qubits.
We show a simple application for four-qubit RGHZ-symmetric states.
It is indicated that there exist four-qubit states in the $G_{abcd}$ class admitting a bilocal model in the form of  \emph{1 versus 3}.
To study the relationship between GME and GMNL in some highly symmetric multipartite states,
under a classification of entanglement beyond the scheme according to the number of parties,
the main difficulty is to derive a complete characterization of their entanglement.
On the other hand,
an open question left in the three-qubit system is that whether both the \emph{W} and GHZ types of entangled states can admit a fully local model.
One can try to  symmetrize the fully local state $\rho_F$ constructed by Bowles \textit{et al.}   \cite{PhysRevLett.116.130401} with $N=3$ into the GHZ-symmetric form, however, the result is a separable state.
The development of  new methods to construct  fully LHV models for GHZ-symmetric states with GME is still necessary.
An alternative approach may be to determine the entanglement class for a fully local state, such as the mentioned $\rho_F$ and $\rho_2(T_1)$ in our results.
The approach of the convex characteristic curve   \cite{PhysRevLett.108.230502,PhysRevA.77.032310}  offers a possible solution to classify the genuine tripartite entanglement by calculating the amount of three-tangle.


 \begin{acknowledgments}
This work was supported by the NSF of China (Grants No. 11675119, No. 11575125, and No. 11105097).
 \end{acknowledgments}

\bibliography{LHVGHZ0606}

\end{document}